\documentclass[12pt, a4paper]{article}
\usepackage[utf8]{inputenc}
\usepackage{amsmath}
\usepackage[sorting=none,style=numeric-comp,url=false]{biblatex}
\usepackage{mathptmx}
\usepackage{commath}
\usepackage{geometry}
\usepackage{graphicx}
\usepackage{mathtools}
\usepackage{setspace}
\usepackage{bm}
\usepackage{etoolbox}
\usepackage{braket}
\usepackage{indentfirst}
\usepackage{caption}
\usepackage{float}
\usepackage{authblk}
\bibliography{bib}
\AtEveryBibitem{
 \clearfield{urlyear}
 \clearfield{urlmonth}
}
\title{Dyonic Taub-NUT-AdS Spaces:\\ Phase Structures of all Horizon Geometries}

\author[a]{Mohamed Tharwat \thanks{oweg@aucegypt.edu}}
\author[b]{Amr AlBarqawy \thanks{amr.albarqawy@sci.asu.edu.eg}}
\author[b,c]{Adel Awad \thanks{a.awad@sci.asu.edu.eg}}
\author[b]{\\ Esraa Elkhateeb\thanks{dr.esraali@sci.asu.edu.eg}}
\affil[a]{\footnotesize \it Department of Physics, School of Sciences and Engineering, American University in Cairo, P.O. Box 74, AUC Avenue New Cairo, Cairo, Egypt}
\affil[b]{\footnotesize \it Department of Physics, Faculty of Science, Ain Shams University, Cairo 11566, Egypt}
\affil[c]{\footnotesize \it Centre for Theoretical Physics, the British University in Egypt, El Sherouk City 11837, Egypt}
\date{}

\begin{document}
\maketitle
\begin{abstract}
 We study phase structures of Lorentzian Dyonic Taub-NUT-AdS spacetimes for different horizon geometries, which are spherical, flat, and hyperbolic. We check the consistency of our extended thermodynamics approach through satisfying the first law, the Gibbs-Duhem relation, and the generalized Smarr's relation. Although we study the phase structure for the three cases, we give special attention to the flat and hyperbolic cases since they are known to show no phase transitions and weren't studied before. Working in a mixed ensemble, we found that the behaviors of the flat and hyperbolic cases are different from those of a charged black hole. In the latter case, a continuous phase transition occurs at high temperatures and pressures, i.e., above the critical point, but in our cases it occurs at low temperatures and pressures, i.e., below the critical point! Generically, the spherical case is characterized by two critical points with continuous phase transition between them.
\end{abstract}
\newpage
\section{Introduction}

In 1951 Taub\cite{taub_empty_1951} found a vacuum axially-symmetric solution of Einstein field equations with two Killing vectors. Later, this solution was studied by Newman, Tumbrino, and Unti \cite{newman_emptyspace_1963} where they extended it beyond its original form. This solution is characterized by a parameter called the nut $n$, in addition to its mass $m$. The nut parameter creates a string-like singular region in this spacetime, which could be chosen to lie along the z-axis. The existence of these strings leaves the spacetime difficult to interpret. To avoid these strings, one can identify the time direction with periodicity, $\beta=8\pi n$. This was suggested by Misner \cite{misner_flatter_2004}, but this condition adds a restriction to the thermodynamics since it relates the horizon radius $r_h$ with the nut parameter $n$. Misner's idea was widely followed in literature \cite{hawking_nut_1999,hawking_gravitational_1999,chamblin_large_1999,hunter_action_1998,mann_complement_2021} and made the Euclidean version of the metric more attractive to work with. 

Recently, several authors considered new thermodynamic treatments of these solutions with Lorentzian signature, where they relaxed the above periodicity condition \cite{hennigar_thermodynamics_2019,bordo_misner_2019,chen_general_2019, Bordo:2019slw,durka_first_2019,awad_dyonic_2023,awad_lorentzian_2022,awad_topological_2020,frodden_first_2022,BallonBordo:2020mcs,wu_thermodynamical_2019}. These new treatments lead to thermodynamics which is not constrained since they relax the above condition which relates $r_{h}$ and $n$. As a result, it added a new term in the form of $\psi_N \, dN$ to the first law! Furthermore, the entropy obtained in these treatments is the area of the horizon. One of the interesting proposals is the one in \cite{hennigar_thermodynamics_2019,bordo_misner_2019} where the authors suggested a nut charge $N$ and a nut potential $\psi_N=\frac{1}{8\pi n}$, where the charge $N$ is a function of $r_{h}$ and the nut parameter $n$. For a brief review of most of these approaches, please see \cite{awad_lorentzian_2022}.

 For the Taub-NUT solution in flat space, it is known that the nut charge $n$ is the dual mass \cite{demianski_combined_1966,dowker_nut_1974}, i.e. if $m$ is considered to be an electric-type charge, $n$ is the dual magnetic type. This is why the Taub-NUT solution is considered to be a gravitational dyon and the Misner string is analogous to the Dirac string! This naturally introduces a conserved charge $n$, which could be used in constructing consistent thermodynamics of these solutions. This was the main idea behind the approach introduced in \cite{awad_lorentzian_2022} where the Dyonic Taub-NUT thermodynamics was studied. Later, this treatment was extended to Anti de Sitter spaces where the thermodynamics and the phase structure of the Dyonic Taub-NUT-AdS with spherical horizon was studied \cite{awad_dyonic_2023}. Here we follow this approach \cite{awad_lorentzian_2022,awad_dyonic_2023} to extend such treatment to study dyonic Taub-NUT-AdS solutions with different horizon geometries, i.e., flat, and hyperbolic horizon geometries. In this work we calculate the thermodynamic quantities, then show the validity of the first law as well as the generalized Smarr's relation. However, the main focus of this work is to study the phase structure of the flat and hyperbolic cases in detail. Both cases show not only first-order phase transitions but critical behaviors as well. 

The connection between $m$ and $n$ can be seen easily through the generalized Komar integral \cite{kastor_enthalpy_2009} which can be can used to calculate the mass of the spacetime 
\begin{equation} M= -\int_{\Sigma} (\star d\xi+2\Lambda\omega)=\sigma_k m, \end{equation} 
where $k=+1$ for the spherical horizon, $k=0$ for the flat horizon, and $k=-1$ for the hyperbolic horizon. $\Sigma$ is the constant-$r$ surface at infinity, $\xi$ is the one-form associated with the time-like Killing vector $\partial_t$, and $\omega$ is a two-form satisfies $d\omega=\star \xi$. An integral over the Hodge dual at infinity yields the conserved nut charge,
\begin{equation}\label{intro}
 \int_{\Sigma} (d\xi-2\Lambda\star\omega)= \pm \sigma_k N_{k}=\sigma_k \,n\,(k+{4n^2/l^2}).
\end{equation} 
Where the $\pm$ sign in the expression is due to the sign of the off-diagonal term in the metric. Notice that in the Minkowski case, or $\Lambda \rightarrow 0$, the charge is proportional to the nut parameter. This is why the nut parameter is considered to be the dual mass. 

There have been some activities recently focusing on studying the phase structure of dyonic Taub-NUT-AdS spacetimes using different treatments and ensembles \cite{xiao_thermodynamical_2021,Bordo:2019slw}. For example, in \cite{Bordo:2019slw} the authors considered the phase transitions in the spherical case within the canonical ensemble. They analyzed the phase structure for the purely electric Taub-NUT-AdS solution. The critical points were obtained perturbatively when the nut parameter is small compared to the electric charge in the canonical ensemble. Their temperature and pressure expressions lead to an approximate evaluation of the critical points. In contrast, our thermodynamic treatment allows for analytical expressions for critical points, which rendered our analysis to cover the whole range of the solution parameters. 

Also, in the previous works \cite{Bordo:2019slw,awad_dyonic_2023,xiao_thermodynamical_2021,abbasvandi_thermodynamics_2021}, the authors studied the phase structure of Taub-NUT Dyonic solutions with spherical horizons only, and none of them studied the flat or hyperbolic cases. This is why we dedicate this work to studying the phase structure of these latter geometries. 
 
We are interested in these cases since their phase transitions were not studied enough in the literature, especially with non-vanishing nut parameter. For the flat horizon dyonic black holes with vanishing nut parameter, the authors in \cite{dutta_dyonic_2013} found no possible phase transitions in this case. However, some literature studies found some phase transitions for this case \cite{plantz_black_nodate}, but they do that in setups significantly different from ours. For example, in \cite{plantz_black_nodate}, it was necessary to add a complex scalar field for a phase transition to take place. Also, in \cite{hennigar_criticality_2017}, the authors showed that phase transitions can only occur with the addition of generalized quasi-topological terms. Lastly, a transition between a black hole and an AdS soliton was discovered for the flat case with a toroidal horizon \cite{surya_phase_2001}. 

In this work, we show that the phase structures of the flat and the hyperbolic cases exhibit first-order as well as continuous phase transitions. Furthermore, we discuss the phase structure of the three geometries for clear comparisons. We show that the shape of the horizon changes the type of thermodynamic phase transition and the region in which it occurs. This leads to different behaviors for the flat and hyperbolic geometries compared to the spherical case. In particular, it has an impact on the number of critical points obtained and whether they can occur for certain values of the electric potentials or not.

The paper is outlined as follows. We first introduce the solutions with different horizon geometries in section 2. Then, we impose some regularity conditions on the vector potential which is worked out in section 3. The calculations of the mass, charges, potentials, and other quantities are carried out in section 4. In sections 5 and 6, we compute the action and use it to check the consistency of our thermodynamics through the first law, Gibbs-Duhem, and Smarr's relations. Our analysis of the phase structure and phase transitions is discussed in section 7, then we present our conclusion in section 8. 

\section{Thermodynamics}

In this section, we explore the thermodynamics of the general dyonic Taub-NUT-AdS spacetime which depends on five main parameters; $m$, the mass parameter, $p$ and $q$ are the magnetic and electric charges at infinity, $n$, the nut parameter, and $l$ is the AdS radius. These spacetimes are characterized by a parameter $k$, which differentiates between three possible horizon geometries \cite{griffiths_taubnut_2010} with $k=-1, 0, 1$ correspond to the hyperbolic, flat, and spherical horizon geometries, respectively. Here and for the rest of this article, we adopt a system of units in which both Newton's constant and the speed of light are equal to unity. The metrics of these spacetimes take the form:
\begin{equation}\label{1}
ds^2=-f(r)\big(dt+2n g_k(x) d\phi\big)^2+ \frac{1}{f(r)}dr^2+(r^2+ n^2)\big(\frac{dx^2}{1-kx^2}+x^2d\phi^2\big),
\end{equation}
Where $f(r)$ and $g(x)$ are given by, 
\begin{equation}
\begin{aligned}
&f(r)=\frac{l^2(p^2+ q^2-2mr+k(r^2- n^2))+r^4+ 6n^2r^2-3n^4}{l^2(r^2+ n^2)},\\[4pt]
&g_k(x)=-\frac{\sqrt{1-kx^2}}{k}+H_k,
\end{aligned}
\end{equation}

where $H_k$ is some constant of integration. The horizon geometry affects $f(r)$ through the term $k(r^2-\Tilde{n}^2)$. For $g_k(x)$ to be regular as $k \rightarrow 0$, it is convenient to set $H_k$ to be
\begin{equation}\label{C1}
H_k=\frac{1}{k}+{C'}_k.
\end{equation}

To work in a more convenient coordinate system, we introduce the following transformations; for $k=1$ we take $x=sin\:\theta$, for $k=0$ we take $x=x$, while for $k=-1$ we take $x=sinh\:\eta$. This leads to the following metrics \cite{Brill:1997mf,Alonso-Alberca:2000zeh},
\begin{equation} \label{metr1}
\begin{aligned}
&ds^2=-f(r)\big(dt+2n(-cos\:\theta+C_1) d\phi\big)^2+ \frac{1}{f(r)}dr^2+(r^2+ n^2)\big(d\theta^2+sin^2\theta d\phi^2\big),&\\[5pt]
&ds^2=-f(r)\big(dt+2n(\frac{x^2}{2}+{C}_0) d\phi\big)^2+ \frac{1}{f(r)}dr^2+(r^2+ n^2)\big(dx^2+x^2d\phi^2\big),&\\[5pt]
&ds^2=-f(r)\big(dt+2n(cosh\:\eta+C_{\scalebox{0.6}{-1}} ) d\phi\big)^2+ \frac{1}{f(r)}dr^2+(r^2+ n^2)\big(d\eta^2+sinh^2\eta d\phi^2\big).&
\end{aligned}
\end{equation}
To fix the above constants we follow Misner \cite{misner_flatter_2004} and calculate $|\nabla t|^2$,
\begin{equation}
|\nabla t|^2=-{1 \over f(r)}+{4 n^2g(x) \over x^2 (r^2+n^2)},
\end{equation}
Notice that for the spherical case, $k=1$, we have singularities at $x=sin\:\theta =0$, or $\theta=0,\pi$. If we try to cancel one of them by choosing some value for $C_1$ we can not cancel the other, leaving what we call a Misner string. In fact, we need to have two patches for removing the Misner string singularities as was suggested in \cite{misner_flatter_2004} which leads to imposing $\beta = 8\pi n$. But since we are going to keep Misner string, or, $\beta \neq 8\pi n$, we set $C_{1}=0$. For the hyperbolic case, $k=-1$, it is enough to choose $C_{-1}=-1$ to regulate this divergence. Also, for the flat case, $k=0$, it is enough to set $C_0=0$. This is why in the last two cases we do not have Misner string and $\beta \neq 8\pi n$.

For the above solution, the mass parameter is given by,
\begin{equation}\label{mass1}
m=\frac{l^2\big(p^2+ q^2+k (r_h^2- n^2)\big)+r_h^4+ 6n^2r_h^2-3n^4}{2r_hl^2}.
\end{equation}

As stated above, we will not impose the periodicity relation $\beta =8\pi n$ to allow for $r_h$ and $n$ to vary independently, which is a vital aspect of our approach. The temperature for this solution is,
\begin{equation}\label{temperature1}
T=\frac{3(r_h^2+n^2)^{2}-l^2(p^2+ q^2-k (r_h^2+n^2))}{4\pi r_h l^2(r_h^{2}+n^2 )}.
\end{equation}

It is important at this point to mention that it is known from the literature \cite{Astefanesei:2004kn} that the Lorentzian Taub-NUT-AdS solutions possess closed time-like curves (CTC's). Therefore, it is important to analyze the two cases we are interested in here. However, before we do so it is important to keep in mind that these two cases are usually studied with compact horizons, as was done in \cite{Mann:1997iz,Mann:1997zn,Brill:1997mf}, where they were called topological AdS black holes.

-- For the planar case, it is more convenient to go to the following coordinate system; $y/l=x\sin{\phi}, \,z/l=x\cos{\phi}, \, t'=t+{2n \over l^2} y z$. This leads to the metric
\begin{equation} ds^2=-f\,(dt'-{2n \over l^2}\, zdy)^2+f^{-1}dr^2+{(r^2+n^2) \over l^2}\,(dy^2+dz^2), \end{equation}
with the $y\,y$-metric component
\begin{equation} g_{yy}={(r^2+n^2) \over l^2}-{4n^2 \over l^4}z^2 f(r).\end{equation}
Thinking of the horizon as a torus with coordinate ranges $y \in [0,a]$, and $z \in [0,b]$, one can show that for every choice of the parameters $[n,l,m,q,p]$ there is always a value for $b=z_{max}$ which leads to a non-negative $g_{yy}$. To show that one can start by noticing that the only term in $f(r)$ which can compete with $r^2+n^2$ is ${r^4 \over l^2(r^2+n^2)}$. Therefore, it is better to do an asymptotic expansion in $r$. This leads to \\
\begin{equation} g_{yy}=r^2\,(1-{4n^2\,z^2 \over l^4})+n^2\,(1-{20n^2z^2 \over l^4})+{8n^2z^2m \over l^2 r}+O(r^{-2}). \end{equation}
From the first two terms to have $g_{yy}>0$, one must have
\begin{equation} \label{frst}
z_{max}< {l^2 \over 2\,n}. \end{equation} 
and
\begin{equation} \label{second}
z_{max}\leq {l^2 \over 2\sqrt{5}\,n}. \end{equation}
Clearly, relation (\ref{second}) is stronger, but one can still have some values of $z_{max}$ a bit larger than the one extracted from condition (\ref{second}) such that $g_{yy}>0$, but it must be less than the value of $z_{max}$ extracted from condition (\ref{frst}).

\begin{figure}
    \centering
    \includegraphics[width=.5\linewidth]{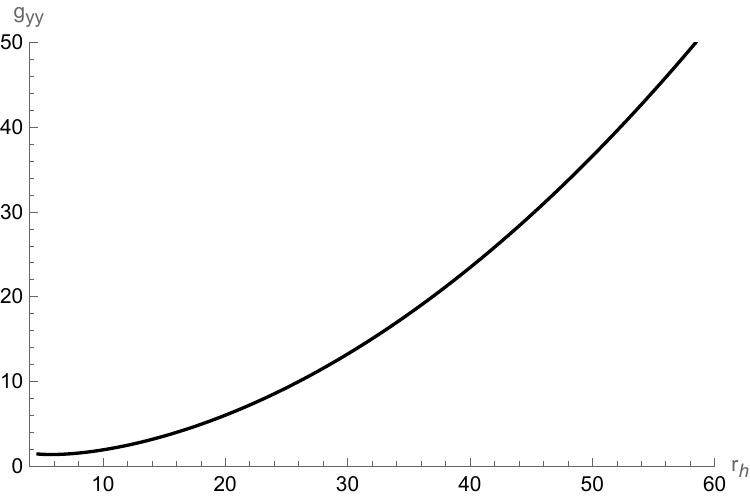}
    \caption{The function $g_{yy}$ showing positive values for all $r_h$. The parameters values are $n = 1,\: l = 4,\: m = 4,\: q = 0.1$,\: p = 0.1, and $z_{max} = 7$.}
    \label{gflt}
\end{figure}

To show the adequacy of this condition we displayed $g_{yy}(r,z_{max})$ as in Eq.~(7), using $[n=1,l=4,m=4,q=0.1,p=0.1, z_{max} = 7]$, see Fig. \ref{gflt}.

\hfill\\
-- For the hyperbolic case, the metric component reads
\begin{equation} \label{gphi}
g_{\phi\phi}={(r^2+n^2)\sinh^2{\eta} }-{4n^2 }(\cosh{\eta}-1)^2 f(r).\end{equation}
For this case one can consider the horizon as a compact surface with a maximum value for $\eta$, or, $\eta\in [0,\eta_{max}]$ as was done in \cite{Mann:1997iz,Brill:1997mf}. Now let us check if we can always have some value for $\eta_{max}$ for every choice of the parameters $[n,l,m,q,p]$, such that $g_{\phi\phi}$ is non-negative. Applying the previous argument and expand $g_{\phi\phi}$ asymptotically in $r$

\begin{equation}
\begin{aligned}
&g_{\phi\phi}=r^2\,\left(\sinh^2{\eta}-{4n^2 \over l^2}(\cosh{\eta}-1)^2 \right)\\[8pt]
&+n^2\,\left(\sinh^2{\eta}-4(\cosh{\eta}-1)^2\,\left({5n^2 \over l^2}-1\right)\right)+{8n^2(\cosh{\eta}-1)^2\,m \over r}+O(r^{-2}).
\end{aligned}
\end{equation}
Again, from the first two terms, one can put conditions on $\eta_{max}$ such that $g_{\phi\phi}>0$. This leads to 
\begin{equation}\tanh^2({\eta_{max}\over 2})< {l^2 \over 4n^2}, \label{3rd} \end{equation}
and
\begin{equation}
\tanh^2({\eta_{max}\over 2})< {l^2 \over 4(5n^2-l^2)}, \label{4th}\end{equation}

We can divide the hyperbolic case into three different cases;
\begin{enumerate}
    \item For $(5n^2-l^2) < n^2$, or $4n^2 < l^2$, condition (\ref{3rd}) is enough to have $g_{\phi\phi}>0$. But in this case, we have $l^2/(4n^2) >1$, which leads to no condition on $\eta_{max}$ since always $tanh^2(x) \leq 1$. 
    \item For $(5n^2-l^2) = n^2$, or $4n^2 = l^2$, the two conditions (\ref{3rd}) and (\ref{4th}) are the same. Again we get $\tanh^2({\eta_{max}\over 2}) <1$, which is always satisfied for any $\eta_{max}$. Therefore, we have no condition for this case on $\eta_{max}$.
    \item For $(5n^2-l^2) > n^2$, or $4n^2 > l^2$, condition (\ref{4th}) is enough to have 
   a positive $g_{\phi\phi}$. Also, since $l^2/(4n^2) <1$, we have ${l^2 \over 4(5n^2-l^2)}<1$, therefore, $\eta_{max}$ is constrained by (\ref{4th}).
    
\end{enumerate} 

\begin{figure}
    \centering
    \includegraphics[width=.5\linewidth]{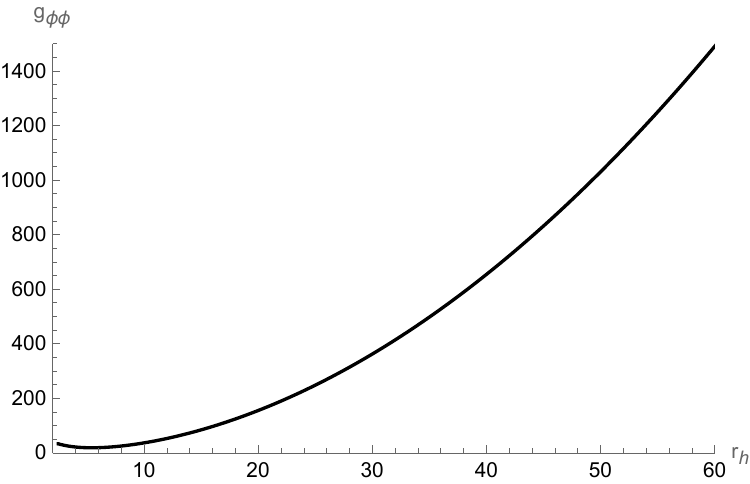}
    \caption{The function $g_{\phi \phi}$ showing positive values for all $r_h$. The parameters values are $n = 1,5\: l = 2,\: m = 4,\: q = 0.1$,\: p = 0.1, and $\eta_{max} = 1.5$.}
    \label{ghyp}
\end{figure}

To show the adequacy of this condition we displayed $g_{\phi \phi}$ as in Eq. (\ref{gphi}) with
$\eta_{max}$ satisfying Eqs. (\ref{3rd}) and (\ref{4th}), using $[n=1.5,l=2,m=4,q=0.1,p=0.1, \eta_{max}=1.5]$, see Fig. \ref{ghyp}.

As a conclusion, for a given set of values $[n,l,m,q,p]$, there are always some choices of the parameters $z_{max}$ and $\eta_{max}$ such that we can avoid the occurrence of CTC's.

\subsection{Gauge potentials}

Here we demand that the gauge potential is well-behaved at the horizon and along the Misner string. These conditions are imposed by the Euclidean path integral \cite{hawking_duality_1995}, which demands the regularity of one-form potential, more precisely, the regularity of its norm at the horizon and along the string. 
The general solution yields the following expressions for the one-form A,
\begin{equation}
\begin{aligned}
&A_t=\frac{(n p+n^2 V-q r+ r^2V)}{(r^2+n^2)},\\[8pt]
&A_\phi= \frac{p\sqrt{1-kx^2}+k L_k}{k}-\frac{2n(np-qr)\big(\frac{\sqrt{1-kx^2}}{k}-H_k\big)}{(r^2+n^2)}.
\end{aligned}
\end{equation}

Where $V$ is the electric potential. Now we need to check the gauge potential to ensure for the different cases. To have a regular $A_\phi$ in the $k=0$ case, $L_k$ should take the form presented in \eqref{C1}, or $L_k=\frac{-p}{k}+{S'}_k$. Now $A_\phi$ takes the form:
\begin{equation}
\begin{aligned}
&[A_\phi]_{\scalebox{0.6}{ k=1}}= p\; cos\:\theta+S_{1}+\frac{(np-qr)\big(-2n\; cos\:\theta
\big)}{(r^2+n^2)},\\[8pt]
&[A_\phi]_{\scalebox{0.6}{ k=0}}= \frac{\big(p(n^2-r^2)-2q\:r\:n\big)x^2+2\big(S_0(r^2+n^2)\big)}{2(r^2+n^2)},\\[8pt]
&[A_\phi]_{\scalebox{0.6}{ k=-1}}=-p\; cosh\:\eta+S_{\scalebox{0.6}{-1}} +\frac{(np-qr)\big(2n\; cosh\:\eta-2n\big)}{(r^2+n^2)}.\end{aligned}
\end{equation}

Enforcing the regularity of $A^2$ at the horizons leads to the following condition 
\begin{equation}\label{11}
q=\frac{np+V(r_h^2+n^2)}{r_h},
\end{equation}
which is the same for all three geometries. After imposing this condition on the norms
\begin{equation}
\begin{aligned}
&[A^2]_{\scalebox{0.6}{ k=1}}= \frac{(cos\: \theta (p+2nV)+S_1)^2}{(r^2+n^2)^2 sin^2\: \theta},\\[8pt]
&[A^2]_{\scalebox{0.6}{ k=0}}= \frac{(2V x^2n-2 S_0+p x^2)^2}{4(r^2+n^2)^2\cdot x^2},\\[8pt]
&[A^2]_{\scalebox{0.6}{ k=-1}}= \frac{(cosh\: \eta (p+2nV)-2n V-S_{\scalebox{0.6}{-1}})^2}{(r^2+n^2)^2 sinh^2\: \eta},
\end{aligned}
\end{equation}
where the denominators vanish for $cos\: \theta = \pm 1$, $cosh\: \eta =1$, and $x^2=0$.
Therefore, regularity demands that

\begin{equation}
\begin{aligned}
 S_1=&\mp (p+2nV),\hspace{0.3 in} 
 S_0=&0, \hspace{0.3 in} 
 S_{\scalebox{0.6}{-1}}=&p. \hspace{0.3 in} 
\end{aligned}
\end{equation}
We chose $S_1$ such that we remove Dirac-like string using two patches. In the other cases it is clear that we have no Dirac-like string.
Notice that the singular behavior along the $\pm z$ axis in the spherical case forces us to use two patches for $A$, these are the patches that leave Dirac string invisible. In contrast, for the flat and the hyperbolic cases, one patch is enough to cover the entire space.

To study the effect of these restrictions on the mass and temperature, we impose the condition on the horizon by substituting for $q$ in their expressions. Relation \eqref{temperature1} now reads,
\begin{equation}\label{temp1}
T=\frac{1}{4 \pi l^2 r_h^3}\left[3r_h^2(n^2+r_h^2)-l^2 (p^2+2 p n V+ V^2 ( n^2+r_h^2)-k r_h^2)\right].
\end{equation}
While the mass parameter \eqref{mass1} becomes,
\begin{equation}\label{mass2}
\begin{aligned}
m =& \frac{1}{2 l^2 r_h^3} \left[r_h^6+\big(6 n^2 + l^2 (k + V^2)\big) r_h^4 \right. \\
&\left. +\big(l^2 (n^2 (V^2-k)
+(n V-p)^2)-3 n^4\big) r_h^2+l^2 n^2 (n V-p)^2\right].
\end{aligned}
\end{equation}

\subsection{Mass and other charges}
In this subsection, we calculate the mass, charges, and potentials. We begin with the mass and nut charge calculations before moving to the electric and magnetic charges and their potentials. To generalize our calculations we use a term $\sigma_k$ to signify the constant scaling term picked up when integrating over a constant-$r$ surface, bearing in mind that it varies for each spacetime. \vspace{0.2em}\\
The mass is calculated through the Komar integral,

\begin{equation}
    - \int_{\partial\Sigma} (\star d\xi +2\Lambda\omega)=\sigma_k m.
\end{equation}

Where $\partial\Sigma$ is the constant $t$ and $r$ surface in the limit $r\rightarrow \infty$, \,$\xi^\mu$ is the time-like killing vector $(\partial_t)$, "$\star$" is the Hodge star operator, and $\omega$ is a two-form such that $d\omega=\star \xi $\cite{kastor_enthalpy_2009}. The nut charge, $N_k$, is the integral of the Hodge dual to the mass over the boundary,
\begin{equation}
    \int_{\partial\Sigma}  (d\xi -2\Lambda\star\omega)=-\sigma_k N_k,
\end{equation}
Where
\begin{equation}
\begin{aligned}
N_{\scalebox{0.8}{$k$=1}}\:=n(1+\frac{4n^2}{l^2}),\:\:\:
N_{\scalebox{0.8}{$k$=0}}\:=\frac{4n^3}{l^2},\:\:\:
N_{\scalebox{0.8}{$k$=-1}}=n(-1+\frac{4n^2}{l^2}).
\end{aligned}
\end{equation} 
A more in-depth derivation for the above quantities was carried out for the spherical case in \cite{awad_dyonic_2023}.

 Now, we focus on calculating the electric and magnetic charges. We use Komar integrals over the boundary to find the total charge within a surface of constant $r$. Defining the two-form $B$ as $dB= \star dA$, the electric and magnetic charges contained inside some closed surface $\partial \Sigma$ are given by
\begin{equation}
\begin{aligned}
&Q_{e}=  \int_{\partial \Sigma} \star F= \int_{\partial \Sigma} dB, \\
&Q_{m}=  -\int_{\partial \Sigma} F= - \int_{\partial \Sigma} dA. \\
\end{aligned}
\end{equation}

The charges at the horizon and radial infinity are
\begin{align}  \label{charges}
    &Q^h_e=\sigma_k \frac{V(r_h^2-n^2)-np}{r_h}, \;\;\;\;\;\;\;
    Q^\infty_e=\sigma_k q, \\
    &Q^h_m=\sigma_k (p+2nV),\;\;\;\;\;\;\;
    Q^\infty_m=\sigma_k p \, .
\end{align}

The potentials $\Phi_e$ and $\Phi_m$ are defined as
\begin{equation}
\begin{aligned}
&\Phi_e=\xi^\mu A_\mu \Big\rvert_\infty-\xi^\mu A_\mu \Big\rvert_{r_h},\:\:\:\:
\Phi_m=\xi^\mu B_\mu \Big\rvert_\infty-\xi^\mu B_\mu \Big\rvert_{r_h}
\end{aligned}
\end{equation}

These potentials are independent of horizon geometries, they are found to be 
\begin{equation}
\begin{aligned}
&\Phi_e=V,
&\Phi_m=\frac{p+nV}{r_h}.
\end{aligned}
\end{equation}

One can write the electric and magnetic charges on the horizon in terms of the charges at infinity, $n$, and the opposite electromagnetic potential. They are given by,
\begin{equation}
\begin{aligned}
&Q^h_e=Q^\infty_e -2\sigma_k n\Phi_m, 
&Q^h_m=Q^\infty_m +2\sigma_k n\Phi_e .
\end{aligned}
\end{equation}

It is interesting to notice that there is no dependence on horizon geometry when it comes to the charges, except for a scaling factor. For all cases, there is a difference between the value of the charge at infinity and the horizon.

It is intriguing to note that the existence of the topological strings does not affect the charges. The flat and hyperbolic cases have no Misner-like strings, yet there are differences between the charges at the horizon and infinity! 

\subsection{Action calculation}
Here we calculate the action of the spacetime, which consists of a bulk term and two boundary terms. It is given by 
\begin{equation}
\begin{aligned}
&I=I_{EM}+I_{GH}+I_{CT},
\end{aligned}
\end{equation}

The first term is the Einstein-Maxwell action, the second is the Gibbons-Hawking boundary term, and the rest are called boundary counterterms \cite{emparan_surface_1999,balasubramanian_stress_1999} which are used to regularize actions in AdS spaces. These are given by, 
\begin{equation}\label{action1}
\begin{aligned}
&I_{EM}= \frac{-1}{16\pi }\int_M d^4x \sqrt{-g}\Big(R+\frac{6}{l^2}-F^2\Big), \\[8pt]
&I_{GH}= \frac{-1}{8\pi }\int_{\partial M} d^3x\sqrt{-h}\:K, \\[8pt]
&I_{CT}= \frac{1}{4\pi l }\int_{\partial M}d^3x\sqrt{-h}\Big(1+\frac{l^2}{4}R\Big).
\end{aligned}
\end{equation}

Using the equations provided above, we calculate the action for each spacetime. The action takes the form

\begin{equation}\label{action2}
\begin{aligned}
&I=\frac{\sigma_k\beta}{2l^2r_h}((n V+p)^2+m r_h-V^2 r_h^2) l^2-3 n^2 r_h^2-r_h^4).\\
\end{aligned}
\end{equation}

Where $m$ is given by eqn. (\eqref{mass2}) and the periodicity $\beta$ is given by

\begin{equation}\label{beta}
\beta=\frac{4\pi r_h l^2(r_h^{2}+n^2 )}{3(r_h^2+n^2)^{2}-l^2(p^2+ q^2-k (r_h^2+n^2))}.
\end{equation}
Both quantities are clearly k-dependent.

\subsection{Dyonic Taub-NUT-AdS thermodynamics}
Here we discuss the consistency of our thermodynamic approach through calculating other thermodynamic quantities and checking the first law, Smarr's relation, and the Gibbs-Duhem relation.
\\

Using the action of each spacetime, we can calculate the entropy, which is given by the relation

\begin{equation}
S=\beta \partial_\beta I-I,
\end{equation}

The resulting entropy is the same for all three spacetimes, up to the scaling factor $\sigma_k$, which is the quarter of the horizon area,
\begin{equation}\label{entropy}
S=\sigma_k \pi(r_h^2+n^2).
\end{equation}
This is different from the work of \cite{abbasvandi_thermodynamics_2021}, where they considered the entropy to be the surface area of the horizon in addition to entropy contributions from the Misner strings. 

Here, and in the rest of this work, we are going to follow a well-known thermodynamic approach in which the cosmological constant acts as a pressure, where $P=-\frac{\Lambda}{8\pi}=\frac{3}{8\pi l^2}$ and is known as extended thermodynamics\cite{kastor_enthalpy_2009,kubiznak_p-v_2012}. For the Schwarzschild solution in AdS, the conjugate quantity to the pressure is the volume of the black hole. Here we calculate the volume along with other thermodynamic quantities.

Let us with the Gibbs free energy which is defined as $G(T, \bar{n}, \Phi_e, Q_m, P)=I/\beta$. Where we introduce the variable $\bar{n}=\sigma_k n$. Gibbs energy variation is given by,

\begin{equation}\label{dG}
dG=-SdT+\Phi_m dQ_m^h+\Phi_{\bar{n}}d\bar{n}-Q_e^\infty d\Phi_e+VdP, \\
\end{equation} 

Here the volume is the variation of $G$ with respect to $P$ when all other variables are kept constant. From now on, we will refer to the electric potential as $\Phi_e$ to avoid ambiguity while discussing the volume. From \eqref{dG}, we get
\begin{equation}\label{volume}
 V=\Big(\frac{\partial G}{\partial P}\Big)_{T,Q^h_m,\bar{n},\Phi_e}=\frac{\sigma_k 4 \pi}{3}(r_h^3+3 r_h n^2).
\end{equation}
\hfill

This volume agrees with those found in \cite{ballon_thermodynamics_2019,chen_general_2019}. It also reduces to the Schwarzschild volume when we set $n$ to zero. 
\\ 

Variation of the Gibbs energy with respect to its independent parameters enables us to get the remaining thermodynamic quantities. This is important in checking the consistency of our thermodynamic formulation and is useful in comparing the charges and potentials to the ones obtained using the Komar integrals and the entropy calculated above. 

\begin{equation}
\begin{aligned}
&\Big(\frac{\partial G}{\partial \Phi_e}\Big)_{T,Q^h_m,\bar{n},P}=- Q^\infty_e ,\\[8pt]
& \Big(\frac{\partial G}{\partial Q^h_m}\Big)_{T,P,\bar{n},\Phi_e}=\Phi_m,\\[8pt]
& \Big(\frac{\partial G}{\partial T}\Big)_{\Phi_m,P,\bar{n},\Phi_e}=-S.
\end{aligned}
\end{equation}

It is easy to show that the above calculations agree with our previously calculated ones in equations \eqref{charges} and \eqref{entropy}. We now use the Gibbs energy to calculate the conjugate potential of our charge $\bar{n}$. In this treatment, it is more convenient to use the quantity $n$ or $\bar{n}$ instead of $N_k$\cite{awad_dyonic_2023}. The conjugate potential to $\bar{n}$ takes the form,
\begin{equation}
\Big(\frac{\partial G}{\partial \bar{n}}\Big)_{T,P,\Phi_e,\Phi_m}=\Phi_{\bar{n}},
\end{equation}

\begin{equation}
\Phi_{\bar{n}}=\frac{1}{r_h^3} \left(\big(\frac{\Phi_{e}^{2}}{2}-4\pi P r_h^2\big)n^3+\frac{n}{2}\big(r_h^2(4\pi Pr_h^2+3\Phi_{e}^{2}-k)+Q^{h\:2}_{m}  \big)-Q^h_m \Phi_e(r_h^2+ n^2)\right).
\end{equation}
The internal energy of the spacetime is
\begin{equation}
U=M-\bar{n}\Phi_{\bar{n}}-PV,
\end{equation}
Which can be expressed as:
\begin{equation}
U=\frac{\sigma_k}{3 \: r_h} \left(4\pi Pr_h^4+ \frac{3}{2}(8\pi Pn^2+k+\Phi_{e}^2)r_h+\frac{3}{2}(Q^{h\:2}_{m}-n^2\Phi_{e}^2)\right).
\end{equation}

This can also be obtained from the Gibbs energy\eqref{dG} through a series of Legendre transforms. In its differential form, the internal energy is expressed as:
\begin{equation}
dU=TdS+\Phi_m dQ_m^h+\Phi_{\bar{n}}d\bar{n}+ \Phi_edQ_e^\infty-PdV,
\end{equation}

It is then straightforward to check that the first law holds for our formulation. This is done by checking the following partial derivatives and comparing them with previous expressions, i.e.,
\begin{equation}
\begin{aligned}
&\Big(\frac{\partial U}{\partial S}\Big)_{ Q_{e}^{\infty},Q^{h}_{m},\bar{n},V} = T ,&\Big(\frac{\partial U}{\partial Q^{h}_{m}}\Big)_{ S,Q_{e}^{\infty},\bar{n},V} = \Phi_{m}, \\[10pt]
&\Big(\frac{\partial U}{\partial \bar{n}}\Big)_{ Q_{e}^{\infty},Q^{h}_{m},S,V} = \Phi_{\bar{n}} ,&\Big(\frac{\partial U}{\partial Q_{e}^{\infty}}\Big)_{ S,Q^{h}_{m},\bar{n},V} = \Phi_{e},\\[7pt]
\end{aligned}
\end{equation} 
\begin{equation*}
\Big(\frac{\partial U}{\partial V}\Big)_{Q_{e}^{\infty},Q^h_m,S,\bar{n}}= -P.
\end{equation*} 
Now let us check Smarr relation, which can be obtained using the usual dimensional argument of enthalpy, $H=M-\bar{n}\Phi_{\bar{n}}$, one finds
\begin{equation}
H(S,P,Q_e^\infty,Q_m^h,\bar{n})=2\,(  TS- VP)+Q_e^\infty \Phi_e +Q_m^h \Phi_m+\bar{n}\Phi_{\bar{n}}.
\end{equation}

What remains to check is the Gibbs-Duhem relation, which relates Gibbs energy to the other thermodynamic quantities. In this case, the relation takes the form  \cite{awad_lorentzian_2022}:
 \begin{equation}
G=\frac{I}{\beta}=M-\bar{n}\Phi_{\bar{n}}-TS-Q_e^\infty \Phi_e.\end{equation} We have checked this relation using the calculated action and other quantities which is indeed satisfied. 

\section{Taub-NUT-AdS Phase Structure}
Here we study phase structures of the three spacetimes in extended thermodynamics. It is important to mention that the phase structure of the spherical case has been studied in the extended treatment in the Lorentzian \cite{ballon_thermodynamics_2019,awad_dyonic_2023,xiao_thermodynamical_2021,abbasvandi_thermodynamics_2021} and Euclidean cases in \cite{johnson_thermodynamic_2014,johnson_extended_2014,lee_extended_2014,chamblin_large_1999}. Here we consider a more general spherical case compared to that presented in \cite{awad_dyonic_2023}. However, the main aim of this work is to study the phase structures of the flat and hyperbolic cases and highlight the differences between these different horizon geometries.

The spherical case in this work differs from previous works in more than one way. For example, the approach in \cite{ballon_thermodynamics_2019} differs from ours in choosing the thermodynamic variable associated with the nut charges as well as the choice of the ensemble, while in \cite{abbasvandi_thermodynamics_2021}, the entropy of spacetime is not the horizon area. In fact, all previous works covering the Lorentzian Taub-NUT-AdS phase structure were not able to obtain analytic expressions for the critical points \cite{ballon_thermodynamics_2019,xiao_thermodynamical_2021,abbasvandi_thermodynamics_2021}.

The phase structures of the flat and hyperbolic cases were not studied before. We show that both exhibit critical behaviors at some temperature. For the flat geometry case previous works showed that phase transitions rely on the addition of generalized quasi-topological terms \cite{hennigar_criticality_2017}, scalar hair \cite{plantz_black_nodate}, or in some soliton' background \cite{surya_phase_2001} to exist \cite{dutta_dyonic_2013,plantz_black_nodate}. Also, our results are new for the Taub-NUT-AdS dyonic solutions with hyperbolic horizon geometry which exhibits critical behavior as well.

We organize this section as follows: in subsection \ref{crit} we discuss the general conditions for the existence of critical points. In subsection \ref{phstr}, we analyze phase structures of different horizon geometries and discuss their features. 

\subsection{Critical points} \label{crit}
In simple terms, a critical point is the point that separates a first-order and a continuous phase transitions. For continuous phase transitions, we have only one black hole phase, as opposed to the first-order transition where we have two distinct black hole phases. The critical point is the point where the first and second derivatives of the pressure with respect to $r_h$ vanish.

The equation of state in terms of our thermodynamic variables and $r_{h}$ can be obtained from \eqref{temp1}, it takes the form 
\begin{equation}\label{Pstate}
P=\frac{Tr_{h}}{2 (n^2+r_{h}^2)}+\frac{(Q_{m}^{h}-n \Phi_e)^2+(\Phi_{e}^{2}-k)r_{h}^2}{8 \pi r_{h}^2 (n^2+r_{h}^2)},
\end{equation}
Critical points can be found by solving:
\begin{equation} \label{drivp}
\frac{\partial P}{\partial r_{h}}=\frac{\partial^2 P}{\partial r_{h}^2}=0.
\end{equation}
To organize different critical points we classify them as follows:

\subsubsection{Case I: $\Phi_e^2 \ne k$}
This is the general case where the critical values are expressed as,
\begin{equation} \label{cpsne}
\begin{aligned}
&r_{c\pm}=\sqrt{\frac{3(Q_{m}^{h}- n \Phi_e)^2\pm \sqrt{3(Q_{m}^{h}- n \Phi_e)^2\left[3(Q_{m}^{h}- n \Phi_e)^2- n^2 (k-\Phi_e^2)\right]}}{(k-\Phi_e^2)}},\\[15pt]
&T_{c\pm}=\frac{2}{3 \pi n^4} \sqrt{\frac{3 (Q_{m}^{h}-n \Phi_e )^2 \pm \sqrt{3(Q_{m}^{h}-n \Phi_e )^2 \left[3(Q_{m}^{h}- n \Phi_e)^2 - n^2 (k-\Phi_e^2)\right]}}{(k-\Phi_e ^2)}} \times \\
& \;\;\;\;\;\;\; \left(6(Q_{m}^{h}- n \Phi_e)^2 - n^2 (k-\Phi_e^2) \mp 2 \sqrt{3(Q_{m}^{h}-n \Phi_e )^2 \left[3(Q_{m}^{h}- n \Phi_e)^2 - n^2 (k-\Phi_e^2)\right]}\right),\\[15pt]
&P_{c\pm}=\frac{1}{8 \pi  n^4} \left(6(Q_{m}^{h}- n \Phi_e)^2 - n^2 (k-\Phi_e^2) \mp 2 \sqrt{3(Q_{m}^{h}-n \Phi_e )^2 \left[3(Q_{m}^{h}- n \Phi_e)^2 - n^2 (k-\Phi_e^2)\right]} \right).
\end{aligned}
\end{equation}

It is clear that these critical points exist only under the condition $\Phi_e^2 \neq k$. On the other hand, a basic part affecting their existence is the inner square root, since they must be real, therefore,
\begin{equation} \label{insq}
    3(Q_{m}^{h}- n \Phi_e)^2-n^2 (k-\Phi_e^2) \geq 0.
\end{equation}
Also, satisfying this condition depends on the value of $k$. This distinguishes two cases:
\begin{enumerate}
    \item \textbf{Case \textit{i}: $k=-1\, \& \, 0$} \\
        In this case, the condition (\ref{insq}) is always satisfied.
    \item  \textbf{Case \textit{ii}: $k=+1$} \\
    In this case, things are different if $\Phi_e^2>k$ or $\Phi_e^2<k$;
    \begin{enumerate}
        \item \textbf{$\Phi_e^2>k$}: The condition (\ref{insq}) is always satisfied.
        \item \textbf{$\Phi_e^2<k$}: To satisfy (\ref{insq}) in this case, we must have
        \begin{equation}
            3(Q_{m}^{h}- n \Phi_e)^2 \geq n^2 (k-\Phi_e^2) \, ,
        \end{equation}
        which leads to the condition
        \begin{equation} \label{qmcond}
        \begin{aligned}
            &Q_{m}^{h} \geq n \Phi_e + \frac{n}{\sqrt{3}} \sqrt{\left(k-\Phi_{e}^{2}\right)} \\
            \text{or} \\
            &Q_{m}^{h} \leq n \Phi_e -\frac{n}{\sqrt{3}} \sqrt{\left(k-\Phi_{e}^{2}\right)}
        \end{aligned}
             \end{equation}
    \end{enumerate}
\end{enumerate}
In summary, the inner square root is always real for hyperbolic and flat cases, while for spherical case, it is always real if $\Phi_e^2>k$ and real under the condition (\ref{qmcond}) if $\Phi_e^2<k$.\\

The reality of the outer square root, on the other hand, has the following basic condition
\begin{equation} \label{bsccond}
    Q_{m}^{h} \ne n \Phi_e
\end{equation}
In addition, other conditions arise due to the value of $k$ and its relation to the thermodynamic parameters;
\begin{enumerate}
    \item \textbf{Case \textit{i}: $k=-1\, \& \, 0$}\\
    In this case, the $\pm$ reverts the sign inside this square root due to the negative sign of the denominator. Besides, the first term turns out to be negative. This means that $r_+$ is forbidden, while $r_-$ is real if
    \begin{equation}
        3(Q_{m}^{h}- n \Phi_e)^2 < \sqrt{3(Q_{m}^{h}- n \Phi_e)^2\left[3(Q_{m}^{h}- n \Phi_e)^2- n^2 (k-\Phi_e^2)\right]} \, .
    \end{equation}
    This leads to the condition
    \begin{equation}
        n^2 (k-\Phi_e^2) < 0 \,,
    \end{equation}
    which is consistent with our assumption that $k$ is negative or $0$. Note however that this condition is satisfied only if
    \begin{equation} \label{ncond}
        n \ne 0 \, .
    \end{equation}
    This means that the addition of a nut parameter guarantees the existence of a critical point for hyperbolic and flat spacetimes. Note however that when $\Phi_e=0$, this critical point can exist for the hyperbolic horizon only. \\
    \item  \textbf{Case \textit{ii}: $k=+1$} \\
    For the spherical spacetime, we can distinguish two cases
    \begin{enumerate}
        \item \textbf{$\Phi_e^2>k$}: For which analysis similar to the hyperbolic and flat spaces leads to the same result; $r_+$ is forbidden while $r_-$ is always real if $n \ne 0$. Again, the addition of a nut parameter guarantees the existence of one critical point. \\
        \item \textbf{$\Phi_e^2<k$}: In this case, the term $(k-\Phi_e^2)$ is positive, ensuring the reality of $r_+$ under condition (\ref{qmcond}). Note that this critical point exists even if the nut parameter vanishes. However, if $n$ is zero, $Q_m^h$ must not be zero.\\
        On the other hand, the reality of $r_-$ in this case demands that
        \begin{equation}
        3(Q_{m}^{h}- n \Phi_e)^2 > \sqrt{3(Q_{m}^{h}- n \Phi_e)^2\left[3(Q_{m}^{h}- n \Phi_e)^2- n^2 (k-\Phi_e^2)\right]}
    \end{equation}
    
    This leads us to the result
    \begin{equation}
         \, n^2 (k-\Phi_e^2) > 0
    \end{equation}
    which is consistent with our assumptions that $k$ is positive and $\Phi_e^2<k$, also $n \ne 0$. Consequently, if $\Phi_e^2<k$, the spherical spacetime will exhibit two critical points, one of which is a consequence of the existence of the nut parameter.
    \end{enumerate}
\end{enumerate}
From the above discussion, we conclude that the addition of a nut parameter guarantees the existence of at least one critical point irrespective of the horizon geometry. Another critical point also exists for the spherical horizon if $\Phi_e^2<1$, i.e. $\Phi_e^2<k$, even if $n$ vanishes. However, we must have a non-vanishing $Q_{m}^{h}$ if $n$ is zero.

\subsubsection{Case II: $\Phi_e^2 = k$}
This case is consistent only with flat and spherical spacetimes. We can obtain the critical values in this case by setting $\Phi_e^2 = k$ in the equation of state (\ref{Pstate}) and solving using (\ref{drivp}). In doing so, we get one critical point
\begin{equation} \label{cpspcial}
\begin{aligned}
    &r_c= \frac{n}{\sqrt{2}} ,\\[6pt]
    &T_c=\frac{4\, \sqrt{2}\,(Q_{m}^{h}- n \Phi_e)^2}{\pi n^3} ,\\[6pt]
    &P_c=\frac{3\,(Q_{m}^{h}- n \Phi_e)^2}{2\,\pi n^4}
\end{aligned}
\end{equation}
Note that this critical point exists under the condition (\ref{bsccond}) provided that $n \ne 0$. For flat space, this condition is turned to be $Q_{m}^{h} \ne 0$. \\

\noindent Summing all up, we have the following:
\begin{itemize}
  \item For hyperbolic horizon we get one critical point as a result of the addition of the nut parameter, provided that condition (\ref{bsccond}) is satisfied. This point can exist even if $\Phi_e=0$.
  \item For flat horizon we get one critical point. For $\Phi_e^2 \ne 0$, this point is similar to the one obtained for hyperbolic space. For $\Phi_e^2 = 0$, we also have a critical point, which is a consequence of the addition of the nut parameter, provided that $Q_{m}^{h} \ne 0$.
  \item For spherical horizon, the relation between $\Phi_e^2$ and $k$ is crucial in defining the number of critical points. For $\Phi_e^2>1$, there exists one critical point due to the addition of the nut parameter, provided that (\ref{bsccond}) is satisfied. If $\Phi_e^2<1$, two critical points exist. One of these points is due to the addition of the nut parameter but the condition (\ref{bsccond}) must be satisfied. The other exists under condition (\ref{qmcond}), which is not related to the nut parameter. In fact, it exists even if the nut parameter vanishes as long as $Q_m^h$ is non-zero, which is the familiar critical point of the charged AdS solution. Finally, if $\Phi_e^2 = 1$, one critical point exists as a consequence of the nut parameter, provided that condition (\ref{bsccond}) is satisfied.
\end{itemize}

\subsection{ Phase structure} \label{phstr}
 To investigate the phase structure of Taub-Nut-AdS solutions, we must study its Gibbs energy  
\begin{equation}
    \begin{split}
        G =& \frac{1}{12r^3}\Big[-8 \pi  P r^6+3 r^4 (k-\Phi_e^2) +3 n^2 (Q_{m}^{h}-n \Phi_e)^2   \\
          & \;\;\;\;\;\;\;\;  -3 r^2\left(8 \pi  n^4 P + n^2 (k-\Phi_e^2) -3 (Q_{m}^{h}-n \Phi_e)^2\right)\Big] \, .
    \end{split}
\end{equation}

Construction of the phase diagrams is accomplished through the application of Maxwell's equal area law \cite{kubiznak_p-v_2012} to obtain the pressure at which the first-order phase transition occurs and the two horizon radii associated with this transition. But since phase transition occurs between two horizon radii sharing the same temperature and the same Gibbs energy, one can use this to obtain the values of pressure and radii as well. Both methods are equivalent and yield the same results, so we implement the latter. Accordingly, we first equate the Gibbs energies and the temperatures for these two radii
\begin{equation}
\begin{aligned}
    &T_{r_s}=T_{r_b}, &G_{r_s}=G_{r_b} \, .
\end{aligned}
\end{equation}
Where $r_b$ is the larger of the two radii where the phase transition happens, while $r_s$ is the smaller one. Upon matching the Gibbs energies and the temperatures, we get the following two equations
\begin{equation} \label{xy1}
\begin{aligned}
    & 24 \pi P x^4 + \left(8 \pi P y^2 - 3(k-\Phi_e^2)\right) x^3 - \left(24 \pi P n^4 + 3 n^2 (k-\Phi_e^2) + 9 (Q_{m}^{h}-n \Phi_e)^2\right) x^2 \\
    &  + 9 (Q_{m}^{h}-n \Phi_e)^2 x + 3 n^2 (Q_{m}^{h}-n \Phi_e)^2 y = 0 \, ,
\end{aligned}
\end{equation}
and 
\begin{equation}\label{xy2}
    8 \pi P x^3 - \left(8 \pi P n^2 - (k-\Phi_e^2)\right) x^2 + 3(Q_{m}^{h}-n \Phi_e)^2 x + (Q_{m}^{h}-n \Phi_e)^2 y^2 = 0 \, .
\end{equation}
Where the variables $x$ and $y$ are introduced to simplify the expressions. They are given by 
\begin{equation}\label{xy}
\begin{aligned}
 &x=r_b\cdot r_s,
 &y=r_b-r_s .
\end{aligned}
\end{equation}
Solving (\ref{xy1}) and (\ref{xy2}) for $x$ and $y$, we can use (\ref{xy}) to get $r_b$ and $r_s$.
The phase diagram can now be obtained by plugging either radius into the equation of state \eqref{Pstate}. \\

\subsubsection{Hyperbolic phase structure}
\begin{figure} [H]
\begin{minipage}{0.48\linewidth}
\captionsetup{width=.9\linewidth}
\centering
\includegraphics[width=0.76\textwidth]{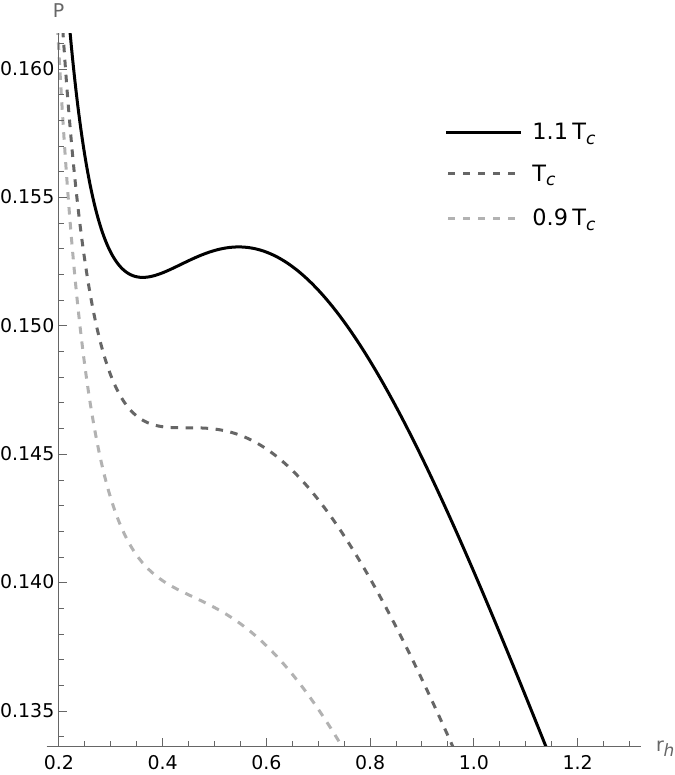}
\caption{Pressure as a function of $r_h$ as $T$ changes around the critical point. $T$ goes from $1.1 T_c$ to $T_c$ to $0.9 T_c$. The critical radius occurs at $r_c=0.446$. The thermodynamic parameters take the following values:
$\Phi_e = 1.1,\: Q_{m}^{h}=1.32,\: n=1,\: T_c=0.347$, and $k=-1$.}\label{hyp-pr}
\end{minipage}
\begin{minipage}{0.48\linewidth}
\captionsetup{width=.9\linewidth}
 \centering
  \includegraphics[width=.76\linewidth]{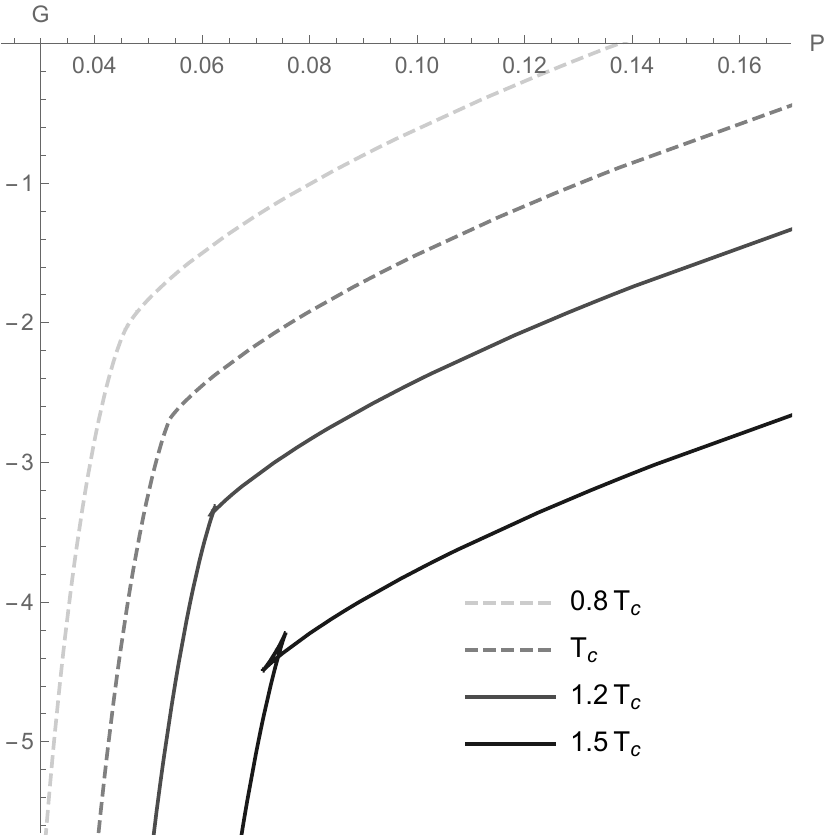}
\caption{Gibbs energy vs pressure for $r\in[0.25,6]$, where
$\Phi_e=0.50 ,\ Q_{m}^{h}=2.00,\: n=1.98,\:k=-1,\:T_{c}=0.353$.}
\label{gibbshyp}
\end{minipage}
\end{figure}

As mentioned above, the hyperbolic spacetime is characterized by a single critical point due to the existence of the nut parameter if $Q_{m}^{h} \ne n \Phi_e$. 
In Fig. \ref{hyp-pr} we plot the pressure as a function of the horizon radius as temperature varies around the critical temperature. There is a first-order phase transition between the small black hole phase (i.e., negative slope region with small radii) and the large black hole phase (i.e., positive slope region with large radii) when the temperature is larger than the critical temperature. As the temperature crosses the critical temperature and decreases further, we get a continuous phase transition between these two phases. \\

For the temperatures above the critical temperature, there are three types of black holes, small, large, and medium (i.e., positive slope region). The medium is unstable since its compressibility is negative, while the small and the large black holes are legitimate phases since their compressibilities are positive. As we decrease the temperature to reach the critical point and below, we get a monotonic function of $r_{h}$ which characterizes the continuous phase transition.

We can see the swallowtail behavior of the Gibbs energy as displayed with respect to the pressure in Fig, \ref{gibbshyp}. The behavior occurs at temperatures above the critical temperature, where the first-order transition occurs from small to large black hole phases. 
\begin{figure}
    \centering
    \includegraphics[width=.5\linewidth]{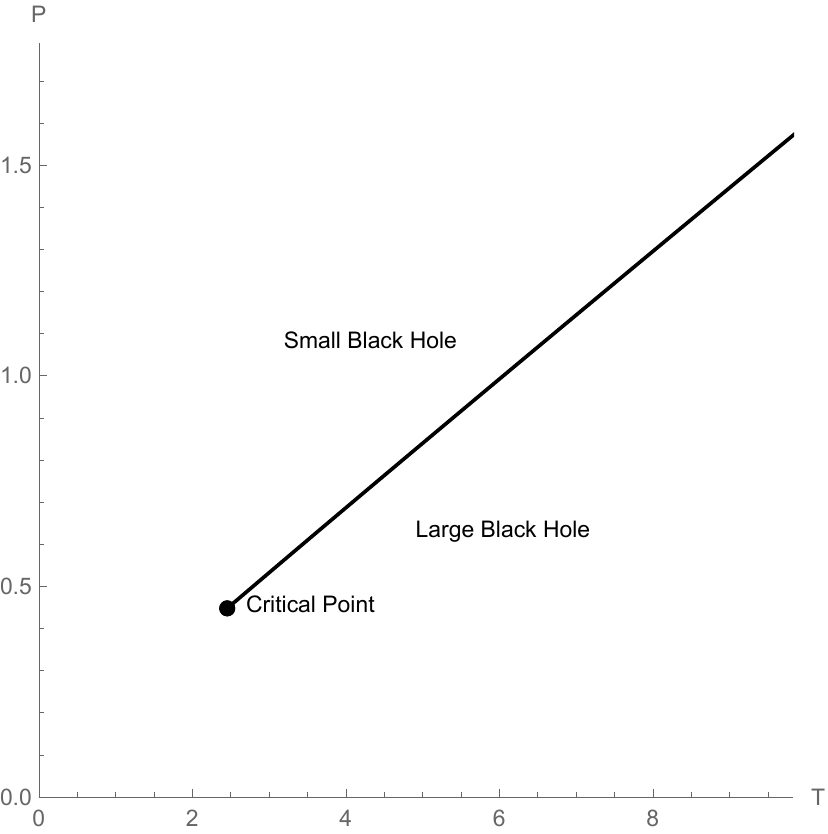}
    \caption{P-T phase diagram for $\Phi_e=0.7 ,\ Q_{m}^{h}=1,\: n=1.50,\:k=-1,\:T_{c}=2.46,\:P_{c}=0.448$.}
    \label{Phase-Structure-Hyperbolic}
\end{figure}

As we mentioned, the first-order phase transition occurs above the critical pressure and temperature, while the continuous phase transition occurs at pressures and temperatures below the critical values. This is contrary to the critical behavior in Van der Waals fluids/charged AdS solutions, where the first-order phase transition occurs below the critical point.
The phase diagram is displayed in Fig. \ref{Phase-Structure-Hyperbolic}. It again shows that a first-order phase transition occurs for temperatures and pressures higher than that of the critical point.

\subsubsection{Flat horizon phase structure}
\begin{figure}[H]
\begin{minipage}{0.48\linewidth}
\captionsetup{width=.9\linewidth}
\centering
\includegraphics[width=0.8\textwidth]{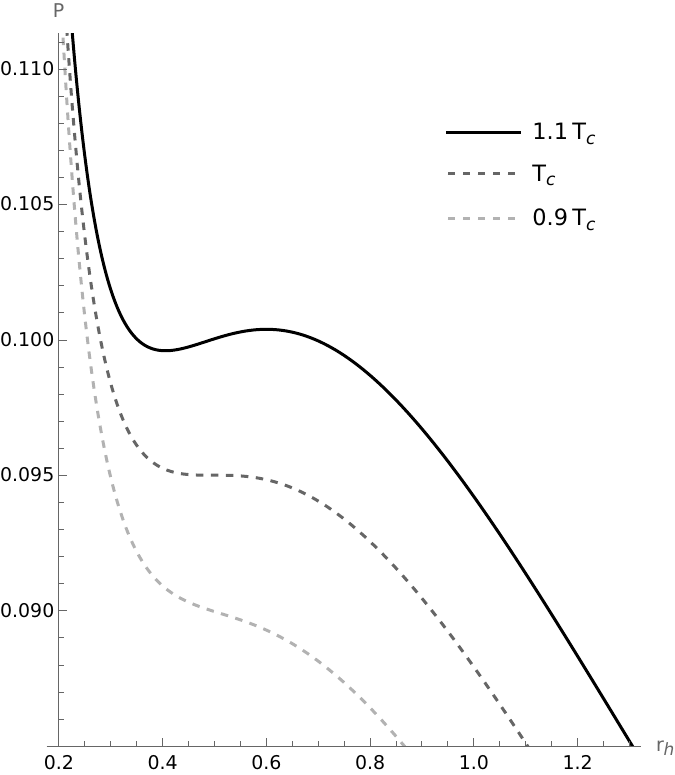}
\caption{Pressure as a function of $r_h$ as $T$ changes around the critical point. $T$ goes from $1.1 T_c$ to $T_c$ to $0.9 T_c$. The critical radius occurs at $r_c=0.496$. The thermodynamic parameters take the following values:
$ \Phi_e = 1.1, \: Q_{m}^{h}=1.32,\: n=1,\: T_c=0.251$, and $k=0$.}
\label{fltpr} 

\end{minipage}
\begin{minipage}{0.48\linewidth}
\captionsetup{width=.9\linewidth}
\centering
\includegraphics[width=0.8\textwidth]{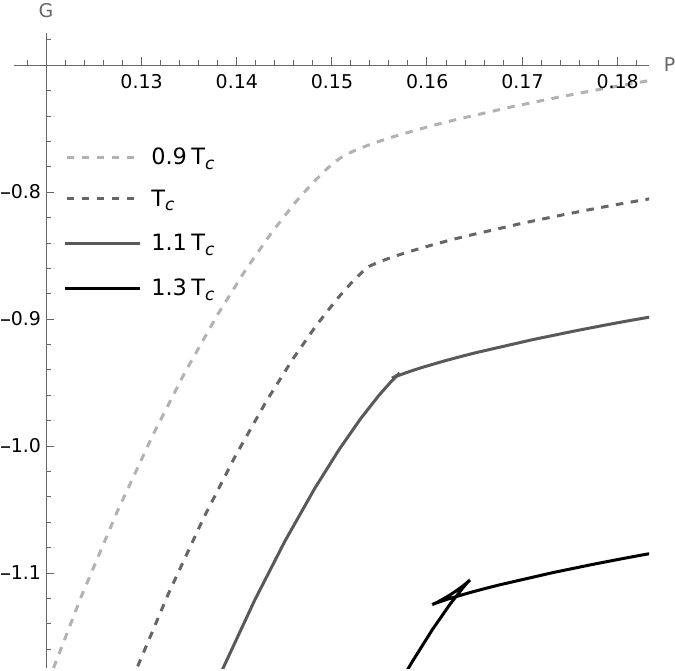}
\caption{Gibbs energy vs pressure for $r\in[0.01,0.3]$, where
$\Phi_e=2.00 ,\ Q_{m}^{h}=2.30,\: n=1.1,\:k=0,\:T_{c}=0.244$.}
\label{gibsflt}
\end{minipage}
\end{figure}

\begin{figure}[H]
\begin{minipage}{0.48\linewidth}
\captionsetup{width=.9\linewidth}
\centering
\includegraphics[width=0.8\textwidth]{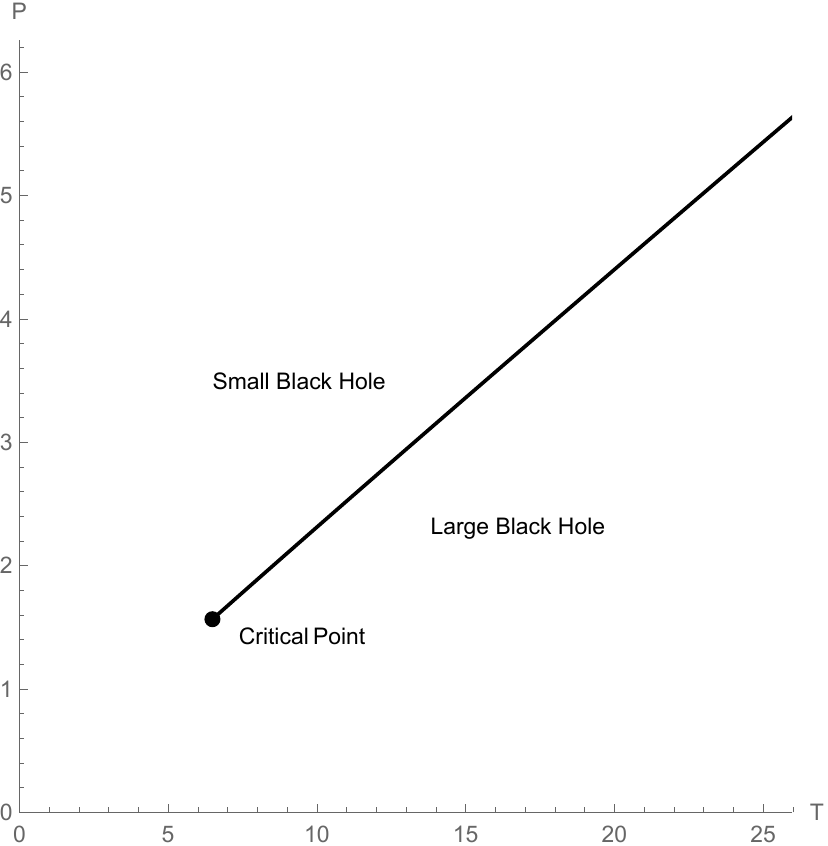}
\caption{P-T phase diagram for
$\Phi_e=0.10 ,\ Q_{m}^{h}=2.30,\: n=1.1,\:k=0,\:T_{c}=6.49,\:P_{c}=1.56$.}
\label{phaseflt}

\end{minipage}
\begin{minipage}{0.48\linewidth}
\captionsetup{width=.9\linewidth}
\centering
\includegraphics[width=0.8\textwidth]{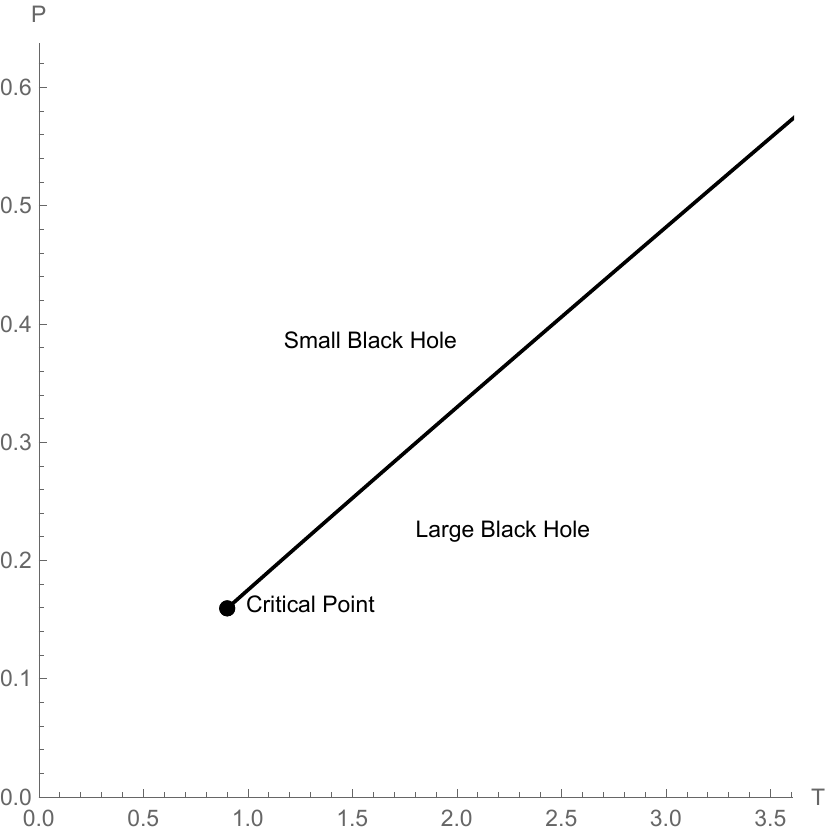}
\caption{P-T phase diagram for
$\Phi_e=0 ,\ Q_{m}^{h}=1.30,\: n=1.50,\:k=0,\:T_{c}=0.902,\:P_{c}=0.159$.}
\label{phsflt0}
\end{minipage}
\end{figure}

As our previous analysis showed, the flat case has a single critical point due to the nut parameter. For $\Phi_e \ne 0$, this point still exist, if $Q_{m}^{h} \ne n \Phi_e$. It resembles the critical point of the hyperbolic case. The first-order transition occurs for temperatures and pressures greater than those of the critical point. This can be seen in Fig. \ref{fltpr}, where we displayed the $P-r_h$ relation for this geometry. This can be seen also in Fig. \ref{gibsflt} for the free energy where the swallowtail behavior of the energy appears at temperatures higher than the critical temperature, ensuring the first-order transition from the small to the large black hole. The phase diagram for this geometry is shown in Fig. \ref{phaseflt}.\\

For $\Phi_e = 0$, we also have a single critical point if $n \ne 0$ provided that $Q_{m}^{h} \ne 0$. This point has the same features as the previous one. The phase diagram for this point is displayed in Fig. \ref{phsflt0}.

\subsubsection{Spherical phase structure}
The spherical case phase structure is richer than the other cases since the values of its parameters affect the number of critical points. Although the behavior of other geometries persists for the spherical case whenever $\Phi_e^2>1$ that its phase structure is indistinguishable from that of the flat and hyperbolic geometries, things are different if $\Phi_e^2<1$. 

\begin{figure}[H]
\begin{minipage}{0.48\linewidth}
 \centering
 \captionsetup{width=.9\linewidth}
 \includegraphics[width=.9
\linewidth]{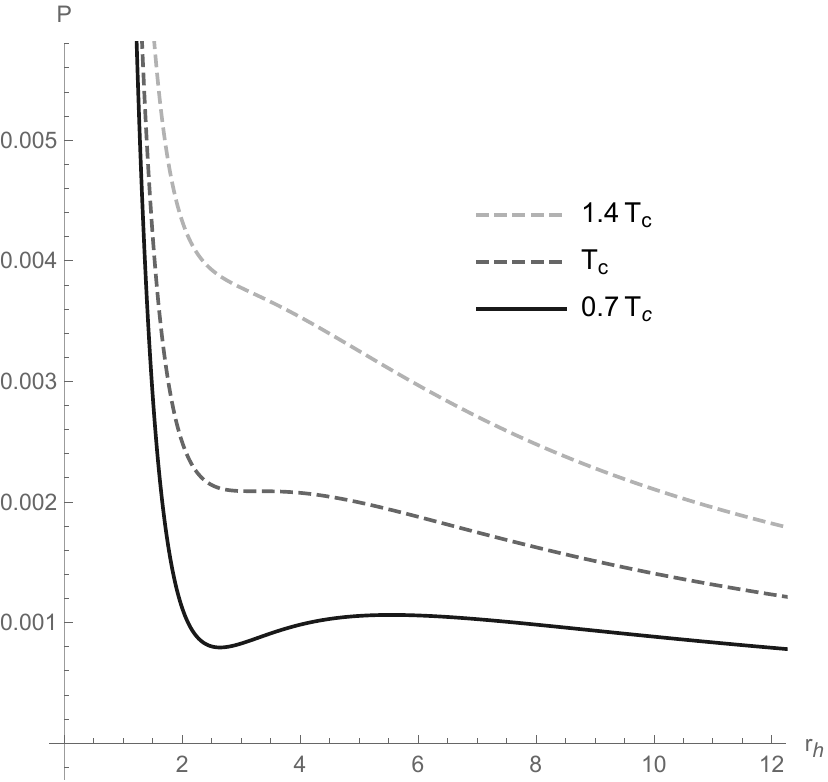}
\caption{Pressure as a function of $r_h$ as $T$ changes around the lower critical point for the spherical horizon when $\Phi_e^2<1$. The first-order transition occurs for temperatures $< T_c$. The thermodynamic parameters are
$\Phi_e=0.50 ,\: Q_{m}^{h}=2.00,\: n=1.98,\:T_{c}=0.0363$, and $k=1$.}
\label{spherical-lower}
\end{minipage}
\begin{minipage}{0.48\linewidth}
\captionsetup{width=.9\linewidth}
 \centering
 \includegraphics[width=.9\linewidth]{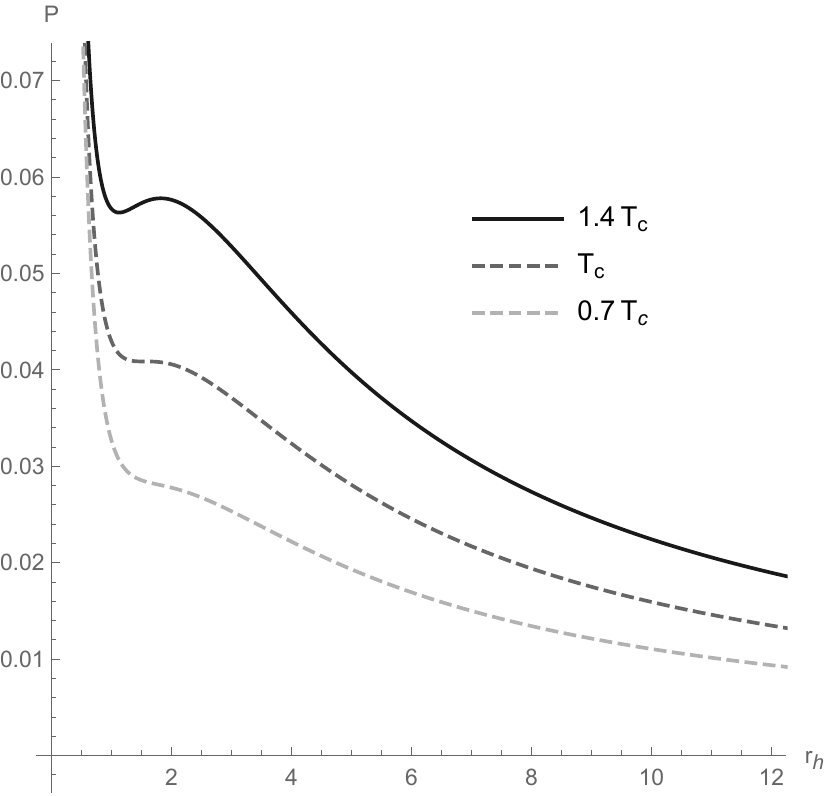}
\caption{Pressure as a function of $r_h$ as $T$ changes around the higher critical point for the spherical horizon when $\Phi_e^2<1$. The first-order transition occurs for temperatures $> T_c$. The thermodynamic parameters are
$\Phi_e=0.50 ,\: Q_{m}^{h}=2.00,\: n=1.98,\:T_{c}=0.127$, and $k=1$.}
\label{Spherical-Higher}
\end{minipage}
\end{figure}

For $\Phi_e^2<1$, there exist two critical points. In addition to the point that exists in all geometries due to the addition of the nut parameter, another one occurs at a lower temperature and pressure. This point is genuine for the spherical horizon whenever $\Phi_e^2<1$ provided that (\ref{qmcond}) is satisfied. This point is not related to the non-vanishing nut parameter. Instead, it occurs even if $n = 0$. However, if $n = 0$, $Q_{m}^{h}$ and $\Phi_e$ must not be zero. In fact, this is the familiar critical point obtained in the charged AdS solution\cite{Chamblin:1999tk,kubiznak_p-v_2012}, where the first-order transition occurs above the critical temperatures and pressures. Figs \ref{spherical-lower} and \ref{Spherical-Higher} show the behavior of the pressure with respect to the horizon radius as the temperature varies around the two critical points.

\begin{figure}[H]
 \centering
 \includegraphics[width=.5\linewidth]{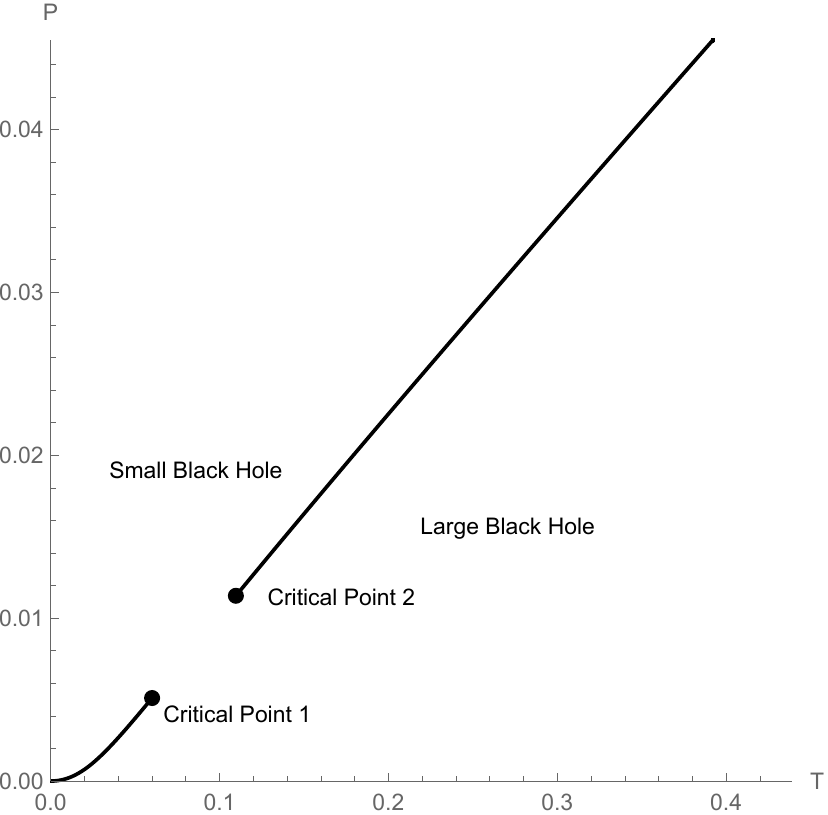}
\caption{P-T phase diagram for the spherical horizon when $\Phi_e^2<1$. The thermodynamic parameters are
$\Phi_e=0.50 ,\ Q_{m}^{h}=2.00,\: n=1.98,\:k=1$.}\label{phase2}
\end{figure}

As shown in Figs \ref{spherical-lower} and \ref{Spherical-Higher}, the first-order phase transition occurs below the critical temperature of the first point and above the critical temperature of the second one. The corresponding phase diagram is presented in Fig. \ref{phase2}.

The phase diagram manifests the existence of two different critical points. Below the first critical point, a first-order phase transition occurs, while Above it a continuous phase transition occurs. Above the second critical point, the transition becomes a first-order again.
When $\Phi_e$ is zero, we get only one critical point which depends on the nut parameter.

\subsubsection*{Special case: Merged points} 
When the inner square root of the critical points, Eq.(\ref{cpsne}), vanishes, the two previous solutions merge into one, in agreement with the results presented in \cite{awad_dyonic_2023}. The whole $P-T$ plane is now divided by a first-order transition curve, except at the critical point as shown in the figure. The condition for merged critical points is 
\begin{equation} \label{mrg1}
    3\left(Q_{m}^{h} - n \Phi_e\right)^2 = n^2 \left(k - \Phi_e^2\right) \, .
\end{equation}

Note that this equation is verified only for $\Phi_e^2 < k$, which is consistent only with the spherical case. The above equation has the solution
\begin{equation}
    n = \frac{Q_{m}^{h}}{\Phi_e \pm \sqrt{(1 - \Phi_e^2)/3} }
\end{equation}

\begin{figure}[H]
\begin{minipage}{0.48\linewidth}
 \centering
 \captionsetup{width=.9\linewidth}
 \includegraphics[width=.9
\linewidth]{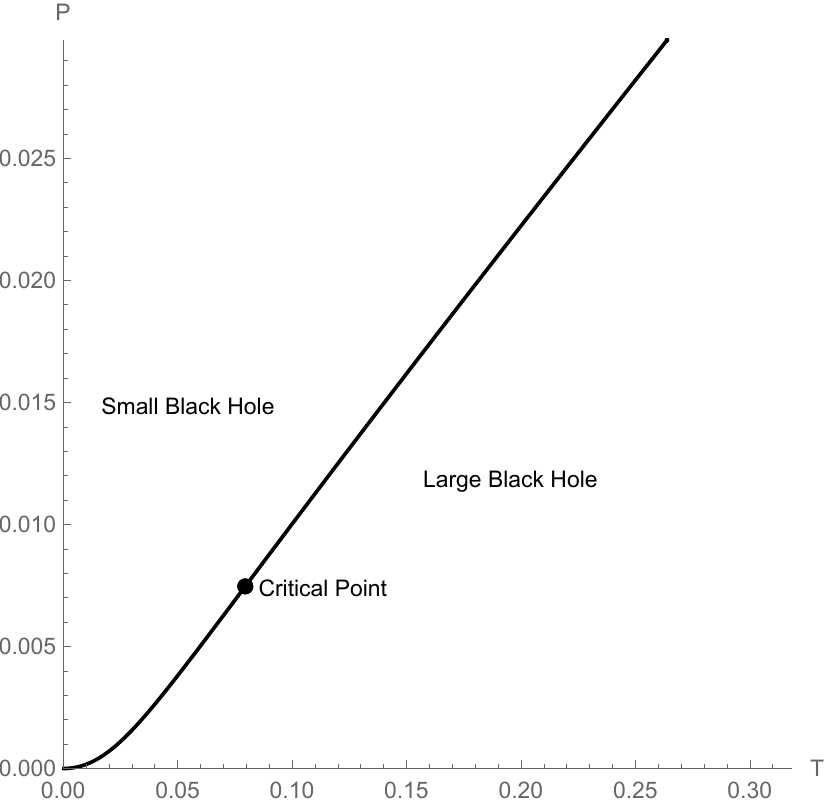}
\caption{P-T phase diagram for
$\Phi_e=0.5 ,\ Q_{m}^{h}=2.00,\: n=2,\:k=1$}\label{spherical-merge}
\end{minipage}
\begin{minipage}{0.48\linewidth}
\captionsetup{width=.9\linewidth}
 \centering
 \includegraphics[width=.9\linewidth]{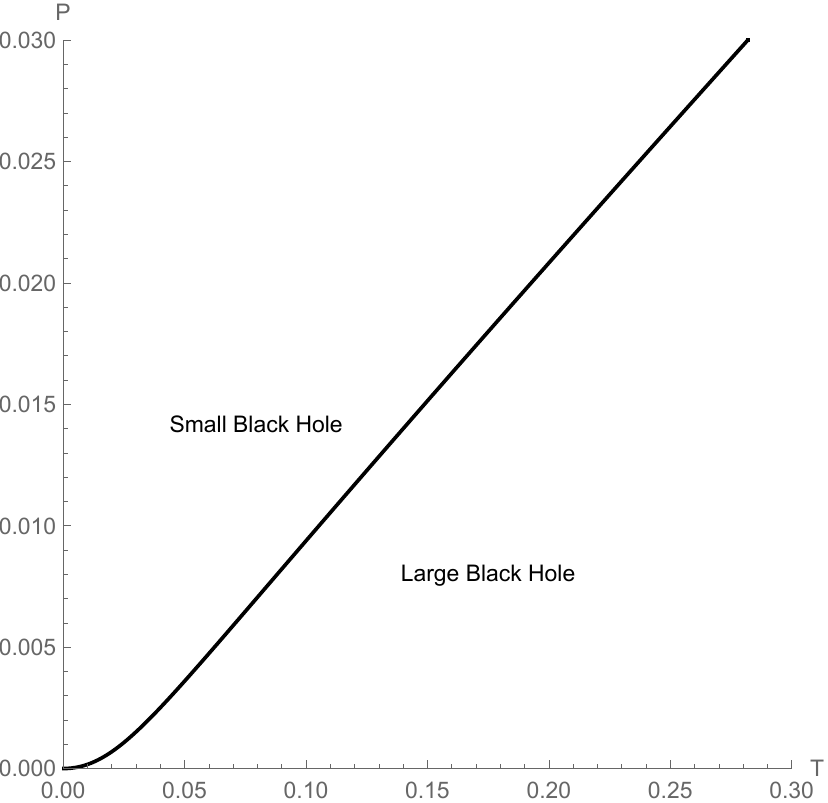}
\caption{P-T phase diagram for
$\Phi_e=0.5 ,\ Q_{m}^{h}=2.00,\: n=2.1,\:k=1$.}\label{spherical-none}
\end{minipage}
\end{figure}

At these values of $n$, we get merged critical points. Fig. \ref{spherical-merge} demonstrates this phenomenon. If $n$ increases further, the critical points disappear. There is no region where a continuous phase transition takes place. The only possible phase transition is now a first order. This is presented in Fig. \ref{spherical-none}. \\

\hfill 

\section{Conclusion}
We used extended thermodynamics treatment to study the phase structure of Lorentzian Dyonic Taub-NUT-AdS metrics with different horizon geometries, i.e., flat, spherical, and hyperbolic. The consistency of this thermodynamics was tested through satisfying the first law, the Gibbs-Duhem, and the Smarr relations.\\

After examining the thermal and mechanical stability of the phases, we found that although we have small, medium, and large black holes, the stable phases are the small and large black holes. In this study, we give more attention to the flat and hyperbolic horizon geometries since their phase structures were not studied before. The flat geometry is of particular interest here since there were no phase transitions reported for it in literature \cite{plantz_black_nodate,dutta_dyonic_2013}.\\ 

The mere existence of a nut parameter introduces a first-order phase transition above a critical temperature and pressure. When the nut parameter vanishes, this critical point vanishes. It is necessary to note that this behavior is different from that of the Van der Waals fluids or charged AdS, since in our cases continuous phase transitions occur at temperatures and pressures below the critical point, while in the Van der Waals case it occurs at temperatures and pressures above the critical point!

Furthermore, the first-order phase transition can occur irrespective of the horizon geometry. This is a novel behavior for the flat and hyperbolic cases. All three cases have the same phase structure when we have $\Phi_n^2>k$.\\

Concerning the spherical case, we have two critical points separated by some region between them. In this region continuous phase transitions occur, confirming the previous work in \cite{awad_dyonic_2023}. The separation between the critical points in the $P-T$ diagram can increase or shrink according to the values of the thermodynamic parameters. For certain values, the critical points overlap, resulting in a single point where a continuous phase transition can take place. Above this point and below it, a first-order transition occurs. When neither critical point exists, there is a first-order phase transition everywhere.\\

There are several avenues in which this work could be expanded. One of them is to study the phase structure of the NUT-AdS solutions in higher dimensions \cite{awad_higher_2006}. Another is to study Kerr-NUT-AdS and Kerr-Newman-NUT-AdS solutions to study the effect of adding angular momentum on possible phases obtained. Lastly, exploring all three dyonic spacetimes in the canonical ensemble would be useful and could yield completely different phase structures.

\newpage
\printbibliography
\end{document}